\documentclass{nato}  %

\usepackage{wasysym} %
\usepackage{amssymb} %
\usepackage{kurt}
\usepackage{makeidx}
\usepackage{bm} %

\usepackage{sprrefn} %

\usepackage[draft]{hyperref}

\newdisplay{guess}{Conjecture}

\makeindex

\begin{document}

\firstpage{1}

\begin{opening}
\title{Zero-Bias Conductance Through Side-Coupled\hfil\break 
Double Quantum Dots }
\author{J. Bon\v ca\email{janez.bonca@ijs.si}}
\runningauthor{J. Bon\v{c}a}
\runningtitle{ Enhanced Conductance ... }
\institute{ J. Stefan Institute, SI-1000 Ljubljana, 
and Department  of Physics, 
FMF, University of Ljubljana, SI-1000 Ljubljana, Slovenia }
\author{R. \v Zitko\email{rok.zitko@ijs.si}}
\institute{ J. Stefan Institute, SI-1000 Ljubljana,Slovenia }

\begin{abstract}
Low temperature zero-bias conductance \index{zero-bias conductance}
through two side-coupled quantum dots is investigated using Wilson's
numerical renormalization group technique.  A low-temperature phase
diagram is computed.  Near the particle-hole symmetric point localized
electrons form a spin-singlet associated with weak conductance. For
weak inter-dot coupling we find enhanced conductance due to the
two-stage Kondo effect when two electrons occupy quantum dots
\index{quantum dots}.  When quantum dots are populated with a single
electron, the system enters the Kondo regime \index{Kondo regime} with
enhanced conductance. Analytical expressions for the width of the
Kondo regime and the Kondo temperature in this regime are given.
\end{abstract}

\end{opening}

\section{Introduction}

The interplay between the Kondo effect \index{Kondo effect} (which involves coupling of the
local moment to the conduction electrons) and the magnetic ordering of
the moments has been studied in the context of bulk materials such as
heavy-fermion metals \cite{jonesvarma1}. Recent advances in
nanotechnology have enabled studies of transport through single as
well as coupled quantum dots \index{quantum dots!coupled} where Kondo physics and magnetic
interactions play an important role at low temperatures. A double-dot
system represents the simplest possible generalization of a single-dot
system which has been extensively studied in the past. Recent
experiments demonstrate that an extraordinary control over the
physical properties, such as the intra-dot coupling,
can be achieved in multiple dot systems 
\cite{dqd-expr1,dqd-expr2,dqd-expr3,pcdqd}. This enables direct
experimental investigations of the competition between the Kondo
effect and the exchange interaction between localized moments on the
dots. One manifestation of this competition is a two stage Kondo
effect \cite{qptmultilevel}. Experimentally, it manifests itself as a
sharp drop in the conductance vs. gate voltage \cite{two_stage}.

The Fano resonance \index{Fano resonance} is a characteristic of
noninteracting electrons for which
the shape of the Fano resonance can be analytically
determined. However, the influence of interactions on the appearance
of Fano resonances remains an open question.  It has been recently
observed in experiments on rings with embedded quantum dots
\cite{ringfano} and quantum wires with side-coupled dots
\cite{scfano}.  The interplay between Fano and Kondo resonance was
investigated using equation of motion \cite{bulka1,bulka2} and Slave
boson techniques \cite{sidedouble}.
 
\begin{figure}[htb]
\centering
\includegraphics[totalheight=2cm]{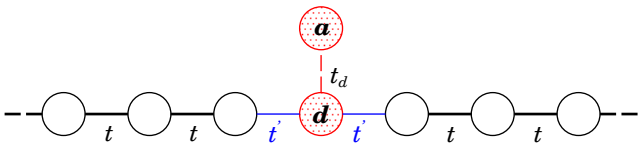}
\caption{Side-coupled configuration of quantum dots.
}
\label{fig0}
\end{figure}

We study a double quantum dot (DQD) in a side-coupled configuration
(Fig.~\ref{fig0}), connected to a single conduction-electron
channel. Similar systems were studied previously with different
techniques, such as: non-crossing approximation \cite{suppression},
embedding technique \cite{topology}, and slave-boson mean field theory
\cite{kang,sidedouble}.  Numerical renormalization group
\index{Renormalization group!numerical} (NRG)
calculations were also performed recently \cite{corn}, where only
narrow regimes of enhanced conductance were found at low temperatures
and no Fano resonances were reported.  We will present a
low-temperature phase diagram of a DQD as a function of intra-dot
coupling strengths and gate-voltage potential indicating regions with
enhanced conductance due to Kondo effect and regions of nearly zero
conductance. In particular, when the intra-dot overlap is large and
DQD occupancy is one, wide regimes of enhanced conductance as a
function of gate-voltage exist at low temperatures due to Kondo
effect, separated by the regimes where localized spins on DQD are
antiferromagnetically (AFM) coupled.  Kondo temperatures $T_K$ follow
a prediction based on the poor-man's scaling and Schrieffer-Wolff
transformation. In the limit when the dot $a$ is only weakly coupled,
the system enters the "two stage" Kondo regime \cite{vojta,corn},
where we again find a wide regime of enhanced conductivity under the
condition that the high- and the low- Kondo temperatures ($T_K$ and
$T_K^0$, respectively) are well separated and the temperature of the
system $T$ is in the interval $T_K^0 \ll T \ll T_K$.

\section{Model and Method}

 The Hamiltonian that we study reads 
\begin{eqnarray} 
H &=&  
\delta (n_d-1) + \delta (n_a-1)
- t_d \sum_\sigma \left( d^\dag_\sigma a_\sigma + a^\dag_\sigma d_\sigma 
\right) \nonumber \\
&+& \frac{U}{2} (n_d-1)^2 + \frac{U}{2} (n_a-1)^2 \\ 
&+& \sum_{k\sigma} \epsilon_k c^\dag_{k\sigma} c_{k\sigma} + 
\sum_{k\sigma} V_d(k) \left( c^\dag_{k\sigma} d_{\sigma} + 
d^\dag_\sigma c_{k\sigma} \right) \nonumber \\
&-& J_{ad}{\bf S}_a \cdot {\bf S}_d, \nonumber
\end{eqnarray} 
where $n_d=\sum_\sigma d^\dag_\sigma d_\sigma$ and  $n_a=\sum_\sigma
a^\dag_\sigma a_\sigma$. Operators $d^\dag_\sigma$ and $a^\dag_\sigma$
are creation operators for an electron with spin $\sigma$ on site $d$
or $a$.  On-site energies of the dots are defined by
$\epsilon=\delta-U/2$. 
For simplicity, we choose the on-site
energies and Coulomb interactions to be equal on both dots.
Coupling between the dots is described by the inter-dot tunnel
coupling $t_d$. Dot $d$ couples to both leads with
equal hopping $t'$.  
Operator $c_{k\sigma}^\dag$ creates a conduction band electron with
momentum $k$, spin $\sigma$ and energy $\epsilon_k=-D \cos{k}$, where
$D=2t$ is the half-bandwidth. Spin operator ${\bf S}=\sum_{s,s'}
c^\dagger_s {\vec \sigma}_{s,s'} c_{s'}$ is defined using Pauli
matrices and $J_{ad}$ represents additional Heisenberg exchange
interaction.  The momentum-dependent hybridization function is
$V_d(k)=-(2/\sqrt{N+1})\, t' \sin{k}$, where $N$ in the normalization
factor is the number of conduction band states.

We use Meir-Wingreen's formula for conductance 
in the case of proportionate coupling \cite{meirwingreen}, which
is known to apply under very general conditions
(for example, the system need not be in a Fermi-liquid 
ground state),
with spectral functions obtained using the NRG technique 
\cite{wilson,sia1,costi,magnetocosti,hofstetter}. 
At zero temperature, the conductance is
\begin{equation} 
G=G_0 \pi \Gamma \rho_d(0), 
\end{equation} 
where $G_0=2e^2/h$, $\rho_d(\omega)$ is the local density of states of 
electrons on site $d$ and $\Gamma/D=(t'/t)^2$.

The NRG technique consists of logarithmic discretization
of the conduction band, 
mapping onto a one-dimensional chain,
and iterative diagonalization of the resulting Hamiltonian \cite{wilson}. 
Only  low-energy part of the spectrum is kept after each iteration step;
in our calculations we kept 1200 states, not counting spin 
degeneracies, using discretization parameter $\Lambda=1.5$. 

\section{Strong intra-dot coupling}

\begin{figure}[htbp] 
\includegraphics[width=6.8cm,angle=-90]{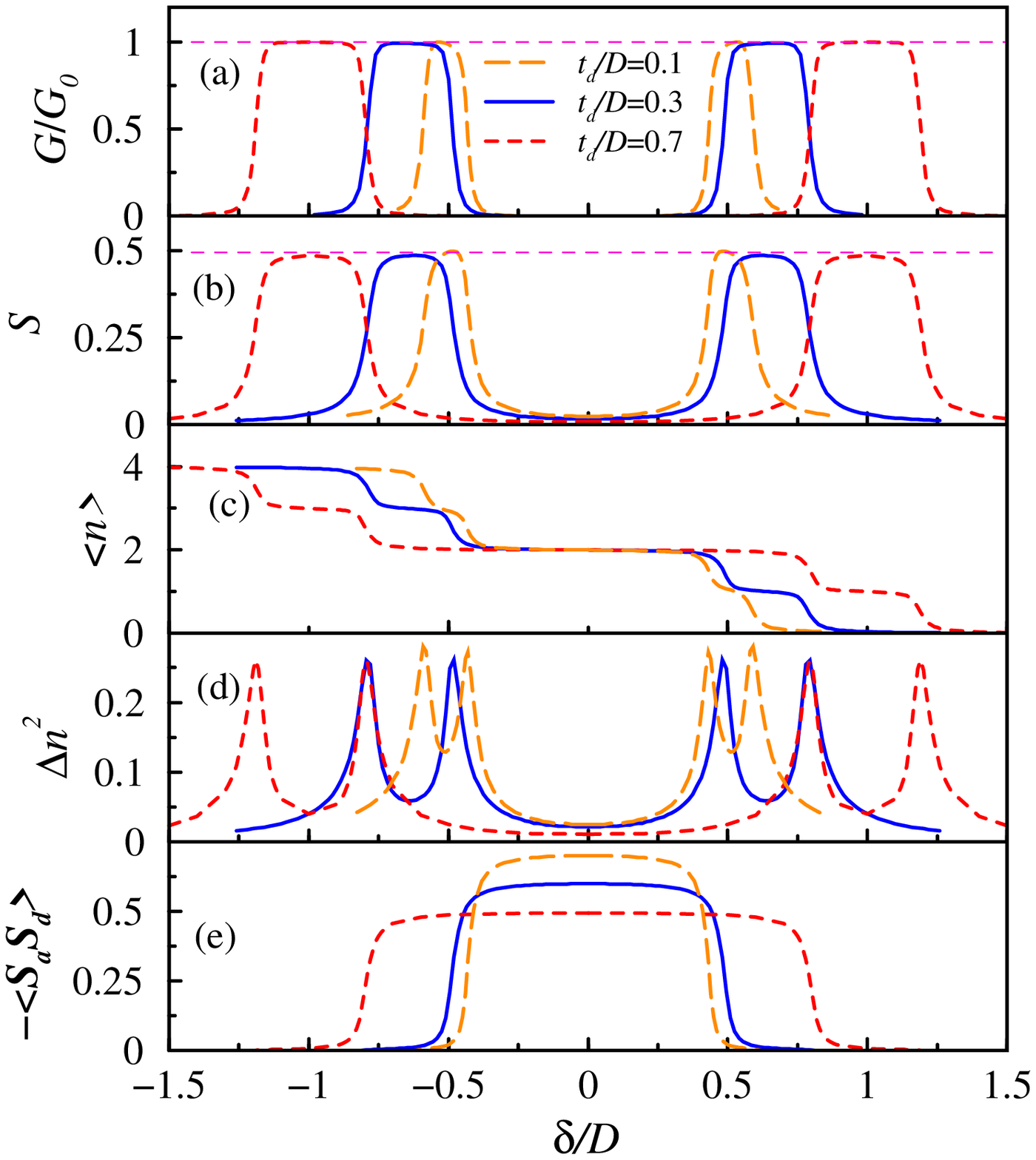} 
\narrowcaption{Conductance and correlation functions of DQD
vs. $\delta$. Besides different values of $t_d$, indicated in the
figure, other parameters of the model are $\Gamma/D=0.03, U/D=1$ and
$J_{ad}=0$. Temperature $T$ is chosen to be far below $T_K$, {i.e.} $T
\ll T_K$.}
\label{fig1}
\end{figure}

\begin{figure}[t] 
\centering
\includegraphics[width=8cm]{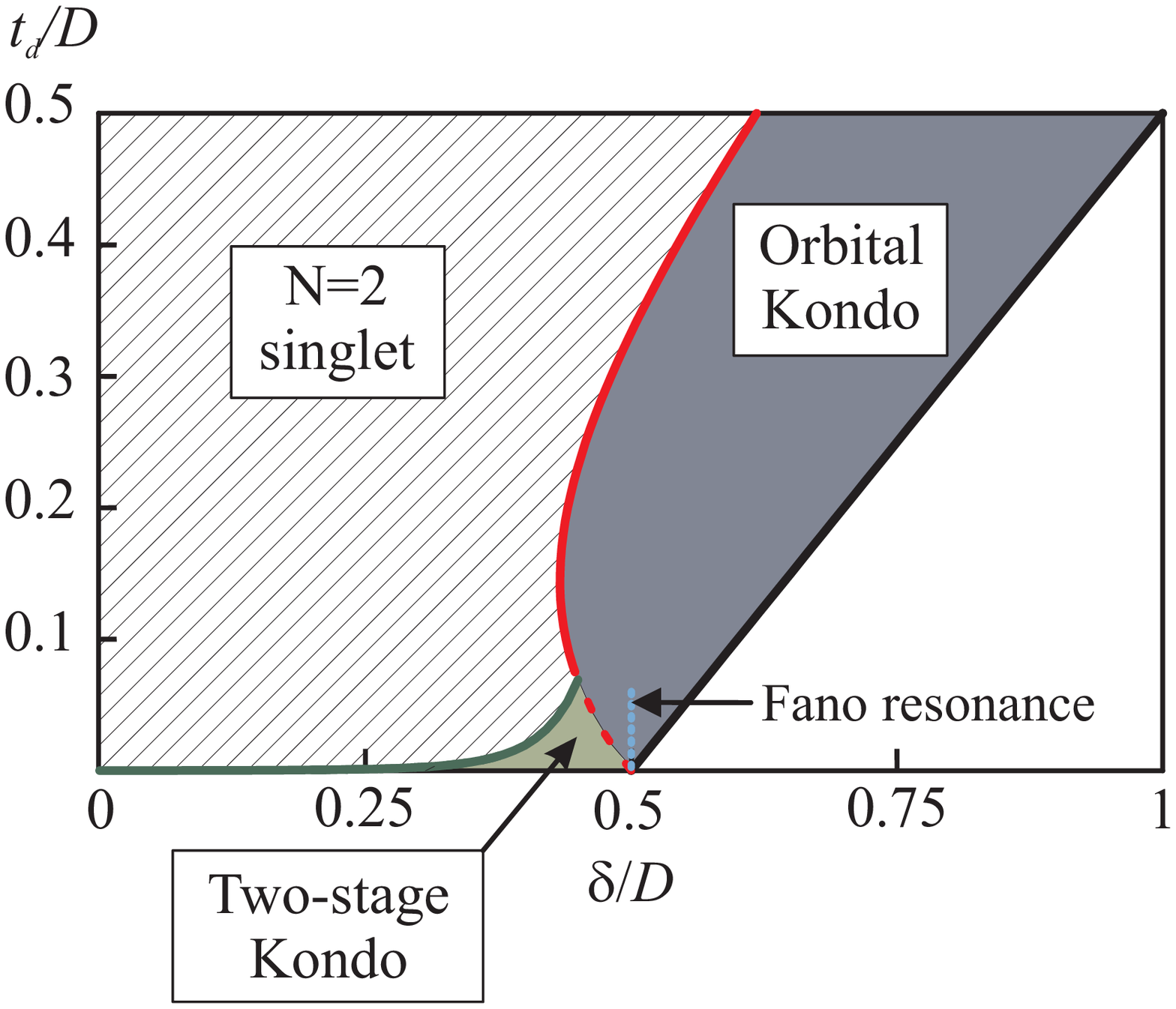}
\vspace{0.1cm}
\caption{ Phase diagram of a DQD for $U/D=1$ and $J_{ad}=0$, obtained
using analytical estimates as given in the text. Grey areas represent
Kondo regimes
\index{Kondo regime} where $S\sim 1/2$, $\langle n\rangle \sim 1$, and
$G/G_0\sim 1$. In shaded area, called spin - singlet regime, where
$S\sim 0$ and $\langle n\rangle \sim 2$, there is enhanced spin-spin
correlation function, {\it i.e.}  $\langle {\bf S}_a \cdot {\bf S}_d
\rangle \lsim -0.5$ and $G/G_0\sim 0$. The two-stage Kondo regime is
explained further in the text.
}
\label{fig2}
\end{figure}

In Fig.~\ref{fig1}a we present conductance through a double quantum
dot at different values of intra-dot couplings vs.  $\delta$.
Due to formation of Kondo correlations, conductance is enhanced,
reaching the unitary limit in a wide range of $\delta$.  Regimes of
enhanced conductance appear in the intervals approximately given by
$\delta_1<|\delta|< \delta_2$, where $\delta_1= t_d
(2\sqrt{1+(U/4t_d)^2}-1)$ and $\delta_2=(U/2+t_d)$. These estimates
were obtained from the lowest energies of states with one and two
electrons on the isolated double quantum dot, {\it i.e.}  $E_1 =
U/2-\delta-t_d$ and $E_2=-2t_d\sqrt{1+(U/4t_d)^2}+U/2$,
respectively. 

Using the above estimates we present a phase diagram in
Fig.~\ref{fig2}. In the gray region, called Kondo regime, with border
lines given by $\delta_1(t_d), \delta_2(t_d)$, the Kondo effect is
responsible for an enhanced conductance.  Kondo plateaus in
Fig.~\ref{fig1} fall in this regime. The Conductance is zero everywhere
outside this region, except in the limit when $t_d\to 0$, where a
two-stage Kondo effect two stage \index{Kondo effect!two-stage} 
is responsible for enhanced conductance, as
discussed further in the text.

To gain further physical insight, we focus on various correlation
functions, defined within the DQD system. In Fig.~\ref{fig1}b we show
$S$, calculated from expectation value $\langle {\bf S}_\mathrm{tot}^2
\rangle=S(S+1)$, where ${\bf S}_\mathrm{tot} = {\bf S}_a+{\bf S}_d$ is
the total spin operator.  $S$ reaches value $1/2$ in the Kondo regime
where $G/G_0=1$. Enhanced conductance is thus followed by the local
moment formation. This is further supported by the average double-dot
occupancy $\langle n\rangle $, where $n=n_a+n_d$, which in the regime
of enhanced conductivity approaches odd-integer values, {i.e.}
$\langle n \rangle=1$ and $3$ (see Fig.~\ref{fig1}c
and~\ref{fig2}). Transitions between regimes of nearly integer
occupancies are rather sharp; they are visible as regions of enhanced
charge fluctuations measured by $\Delta n^2 = \langle n^2\rangle -
\langle n\rangle^2$, as shown in Fig.~\ref{fig1}d. Finally, we show in
Fig.~\ref{fig1}e spin-spin correlation function $\langle {\bf S}_a
\cdot {\bf S}_d \rangle$. Its value is negative between two separated
Kondo regimes where conductance approaches zero, {\it i.e.} for
$-\delta_1<\delta<\delta_1$, otherwise it is zero. This regime further
coincides with $\langle n\rangle\sim 2$. Each dot thus becomes nearly
singly occupied and spins on the two dots form a local singlet (S=0)
due to effective exchange coupling $J_{\mathrm{eff}}=4t_d^2/U$.

\begin{figure}[htbp] 
\includegraphics[width=5.9cm,angle=-90]{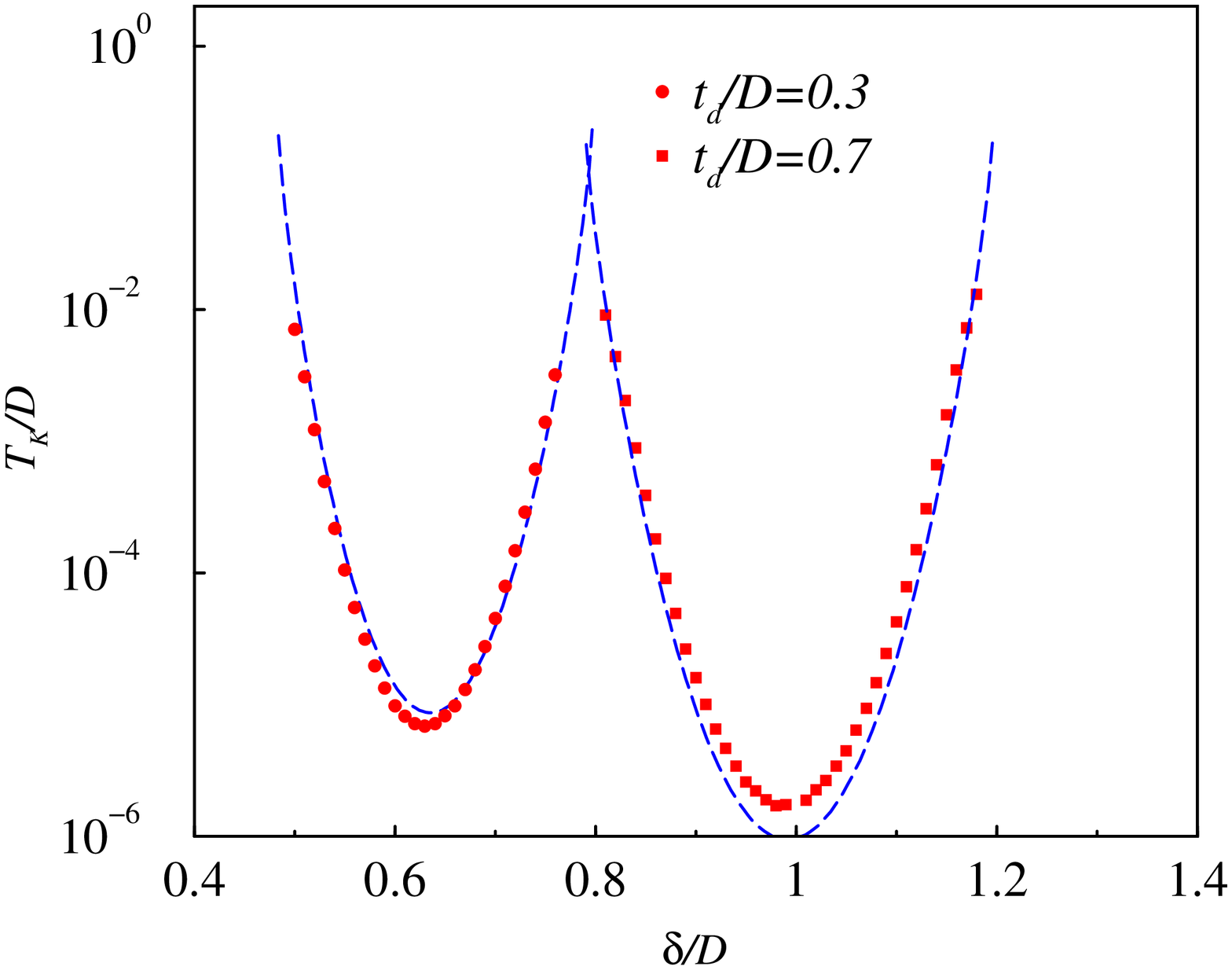}
\narrowcaption{ Kondo temperatures $T_K$ vs. $\delta$ as measured from the
widths of Kondo peaks obtained from NRG calculations (full circles and
squares). Analytical estimate, Eq.~\ref{tk}, is shown using dashed
lines.  The rest of parameters are identical to those in
Fig.~\ref{fig1}.}
\label{fig3}
\end{figure}

In Fig.~\ref{fig3} we present Kondo temperatures $T_K$ vs. $\delta$
extracted from the widths of Kondo peaks. Numerical results in the
regime where $\langle n\rangle\sim 1$ and 3 fit the analytical
expression obtained using the Schrieffer-Wolf transformation that
projects out states with even electron occupancy and
leads to an effective single quantum dot problem with
renormalized parameters. We obtain
\begin{equation} 
T_K=0.182 U \sqrt{\rho_0 J}\exp[-1/\rho_0J] \label{tk} 
\end{equation} 
with $\rho_0 J=\frac{2\Gamma}{\pi}(\alpha/\vert E_1\vert + 
\beta/\vert E_2-E_1\vert)$, where $\alpha=1/2$ and 
\begin{equation} 
\beta = 
\frac{
\left(4 t_d + U + \sqrt{16 t_d^2+U^2}\right)^2
}{
8\left(16 t_d^2+U\left(U+\sqrt{16 t_d^2+U^2}\right)\right)
}.
\label{beta} 
\end{equation} 
The prefactor $0.182U$ in Eq.~\ref{tk} is the effective bandwidth. 
The same effective bandwidth was used to obtain $T_K$ of the Anderson 
model in the regime $U<D$ \cite{sia1,hald}.

\section{Weak intra-dot coupling}

\begin{figure}[htbp] 
\centering
\includegraphics[width=6.5cm,angle=-90]{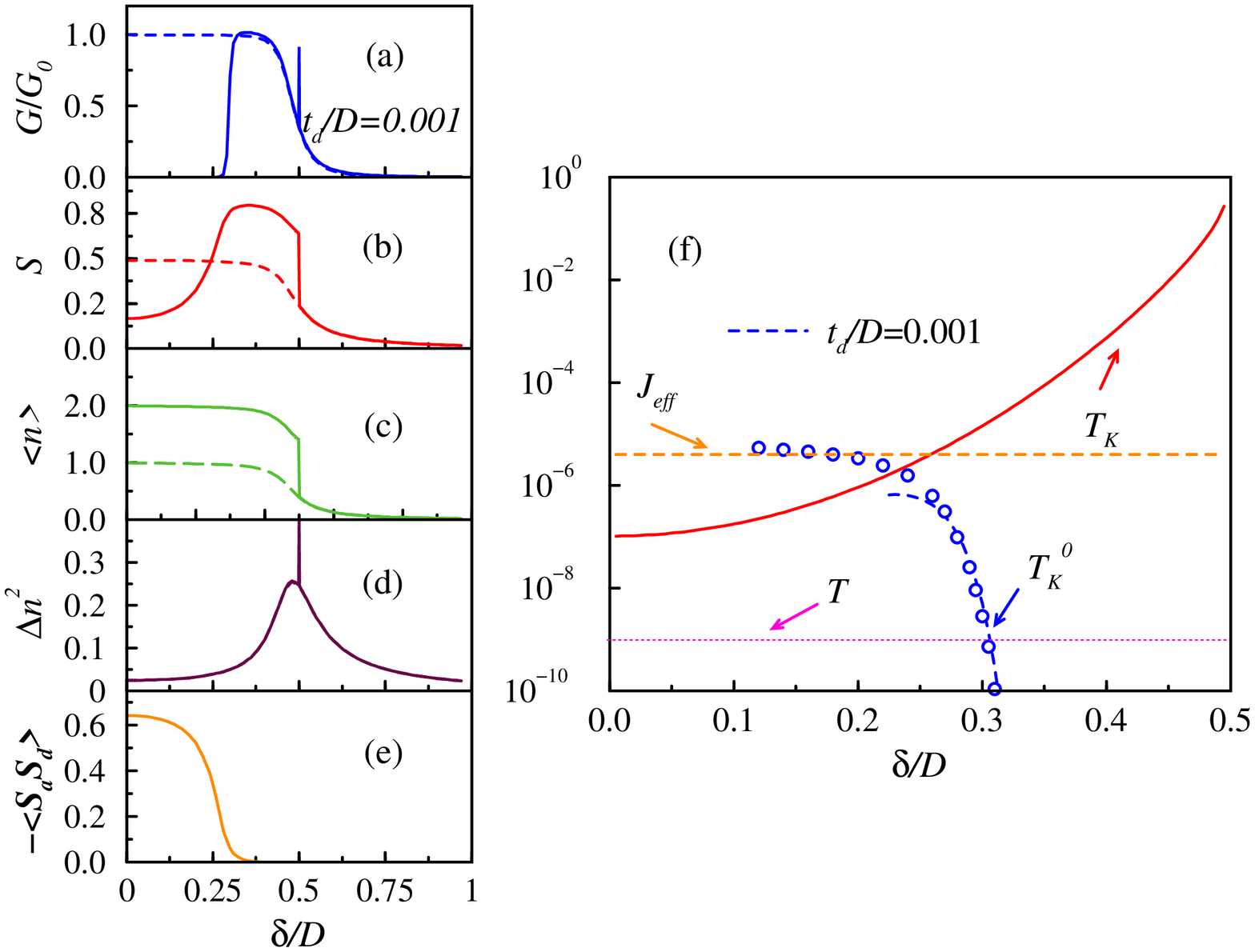}
\caption{Conductance and correlation functions 
at $t_d/D=0.001$ (a,...,e). Dashed lines represent in a) $G/G_0$ and
in b) $S$ of a single quantum dot with otherwise identical
parameters. Dashed line in c) represents $\langle n_d\rangle$ of
DQD. In f) a schematic plot of different temperatures and interactions
is presented as explained in the text. NRG values of the gap in
$\rho_d(\omega)$ at $\omega=0$ and $T\ll T_K^0$ are presented with
open circles.  Values of $J_{\mathrm{eff}}$ and
analytical results of $T_K^0$ are presented with dashed
lines. For analytical estimates of $T_K^0$
$\alpha=2.2$ was used. Other parameters of the model are
$\Gamma/D=0.03, U/D=1$ and $J_{ad}=0$.}
\label{fig5}
\end{figure}

We now turn to the {\it limit when} $t_d\to 0$. Unless otherwise
specified, we choose
the effective temperature $T$ to be finite,
{\it i.e.} $T\sim 10^{-9}D$, since calculations at much lower
temperatures would be experimentally irrelevant.  In this case one
naively expects to obtain essentially identical conductance as in the
single-dot case.  As $\delta$ decreases below $\delta\sim U/2$,
$G/G_0$ indeed follows result obtained for the single-dot case as
shown in Fig.~\ref{fig5}a. In the case of DQD, however, a sharp Fano
resonance appears at $\delta = U/2$. This resonance coincides with the
sudden jump in $S$, $\langle n\rangle$, as well as with the spike in
$\Delta n^2$, as shown in Figs~\ref{fig5}b,c, and d,
respectively. The Fano resonance is a consequence of a sudden charging of
the nearly decoupled dot $a$, as its level $\epsilon$ crosses the
chemical potential of the leads, {\it i.e} at $\epsilon=0$. Meanwhile,
the electron density on the dot $d$ remains a smooth function of
$\delta$, as seen from $\langle n_d\rangle$ in Fig~\ref{fig5}c. With
increasing $t_d$, the width of the resonance increases and at $t_d
\gtrsim 0.1$, the resonance merges with the Kondo plateau and
disappears (see Fig.~\ref{fig1}a). 
The regime where the Fano resonance \index{Fano resonance} exists
is also specified in the phase diagram, Fig.~\ref{fig2}.

We now return to the description of the results presented in Fig.~\ref{fig5}a
in the regime where $\delta<U/2$. As $\delta$  further decreases, the
system enters a regime of the two-stage Kondo effect
\index{Kondo effect!two-stage} 
\cite{corn}. This region is defined by $J_{\mathrm{eff}}<T_K$ (see
also Fig.~\ref{fig5}f), where $T_K$ is the Kondo temperature,
approximately given by the single quantum dot Kondo temperature,
Eq.~\ref{tk} with $\rho_0 J = \frac{2\Gamma}{\pi}(1/\vert \delta-U/2
\vert + 1/\vert \delta+U/2 \vert)$.  This regime is also indicated in
the phase diagram, presented Fig.~\ref{fig3}, where condition
$J_{\mathrm{eff}}=T_K$ was used to separate two-stage Kondo from the
spin-singlet regime.  Just below $\delta<U/2$, $T$ falls in the interval,
given by $T_K^0 \ll T \ll T_K$, where $T_K^0 \sim T_K \exp(-\alpha
T_K/J_{\mathrm{eff}})$ denotes the lower Kondo temperature,
corresponding to the gap in the spectral density $\rho_d(\omega)$ at
$\omega=0$ and $\alpha$ is of the order of one~\cite{corn}.
Note that NRG values of the gap in $\rho_d(\omega)$ (open circles),
calculated at $T\ll T_K^0$, follow analytical results for
$T_K^0(\delta)$ when $J_{\mathrm{eff}}<T_K$, see Fig.~\ref{fig5}f,
while in the opposite regime, {\it i.e.} for $J_{\mathrm{eff}}>T_K$,
they approach $J_{\mathrm{eff}}$.

As shown in Fig.~\ref{fig5}a for $0.3D \lsim \delta <U/2$, $G/G_0$
calculated at $T=10^{-9}D$ follows results obtained in the single 
quantum dot case and approaches value 1. 
The spin quantum number $S$ in
Fig.~\ref{fig5}b reaches the value $S\sim 0.8$, consistent with the
result obtained for a system of two decoupled spin-1/2 particles, where
$\langle {\bf \hat S}^2\rangle=3/2$. This result is also in agreement
with $\langle n\rangle\sim 2$ and the small value of the spin-spin
correlation function $\langle {\bf S}_a \cdot {\bf S}_d \rangle$, presented
in Fig.~\ref{fig5}c and \ref{fig5}e respectively.

With further decreasing of $\delta$, $G/G_0$ suddenly drops to zero at
$\delta\lsim 0.3D$. This sudden drop is approximately given by $T\lsim
T_K^0(\delta)$, see Figs.~\ref{fig5}a and f. At this point the Kondo
hole opens in $\rho_d(\omega)$ at $\omega=0$, which in turn leads to a
drop in the conductivity. The position of this sudden drop in terms of
$\delta$ is rather insensitive to the chosen $T$, as apparent from
Fig.~\ref{fig5}f.

Below $\delta\lsim 0.25D$, which corresponds to the condition
$J_{\mathrm{eff}}\sim T_K(\delta)$, also presented in
Fig.~\ref{fig5}f, the system crosses over from the two stage Kondo
regime to a regime where spins on DQD form a singlet. In this case $S$
decreases and $\langle {\bf S}_a \cdot {\bf S}_d \rangle$ shows strong
anti-ferromagnetic correlations, Figs.~\ref{fig5}b and e.  The lowest
energy scale in the system is $J_{\mathrm{eff}}$, which is supported
by the observation that the size of the gap in $\rho(\omega)$ (open
circles in Figs.~\ref{fig5}f) is approximately given by
$J_{\mathrm{eff}}$. 

\section{Ferromagnetic coupling}

\begin{figure}[htbp]
\includegraphics[width=6.5cm,angle=-90]{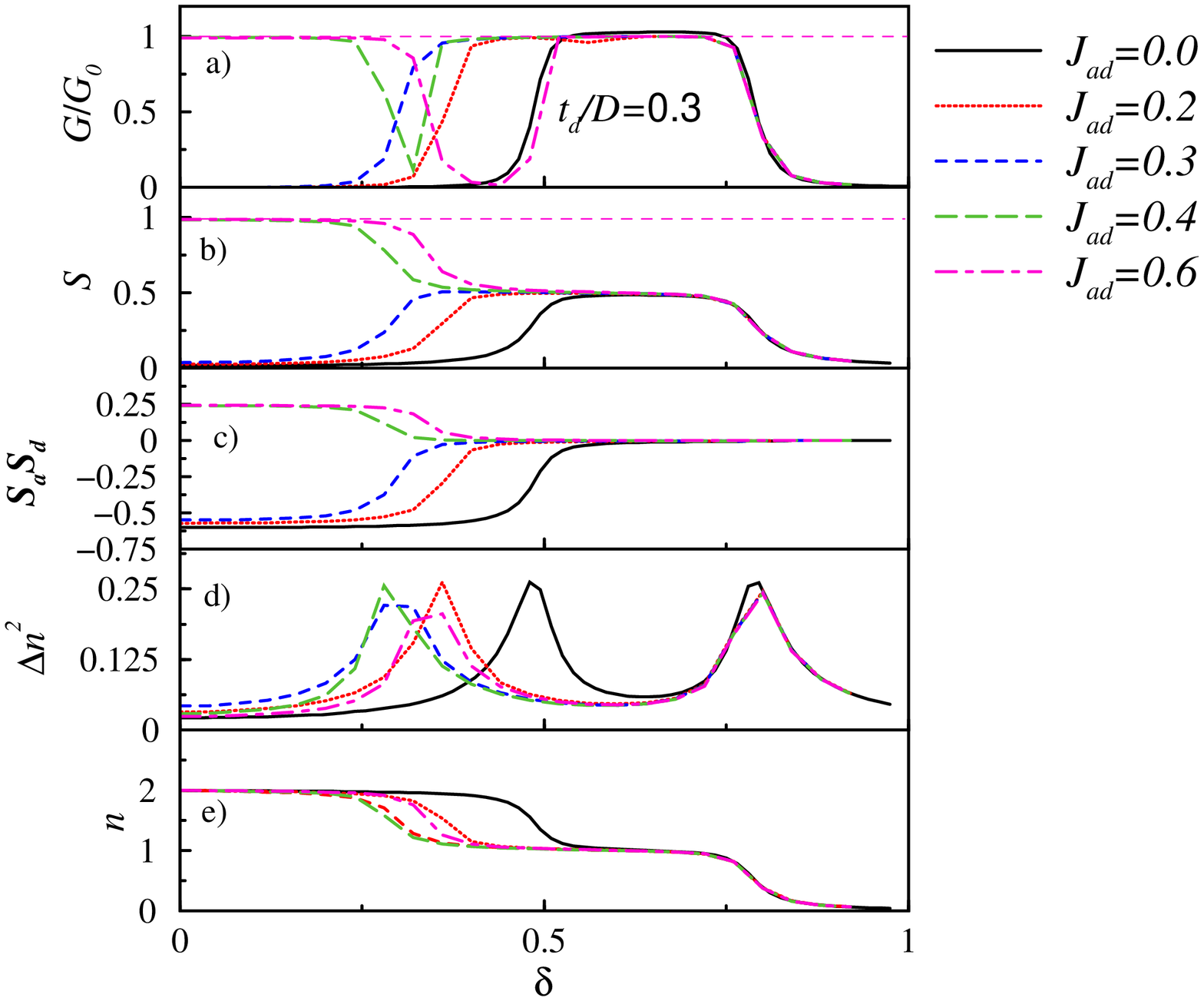}
\narrowcaption{Conductance and correlation functions of DQD vs. $\delta$ for
various values of $J_{ad}$ and $t_d/D=0.3$. Other parameters of the
model are identical to those in Fig.~\ref{fig1}. Temperature $T$ is
chosen to be far below $T_K$, {i.e.} $T \ll T_K$.}
\label{fig6}
\end{figure}

In Fig~\ref{fig6} we present results of conductance and correlation
functions for different values of the ferromagnetic exchange coupling
\index{ferromagnetic coupling}
$J_{ad}$, $t_d/D=0.3$ and $ U/D=1$.  At finite $t_d$ and $J_{ad}=0$
the ground state of an isolated DQD containing two electrons is a
spin singlet state.  At $J_{ad}^c/D=2(\sqrt{4 t_d^2+U^2}-U)\sim 0.33$
spin singlet and triplet states become degenerate. With further
increasing $J_{ad}>J_{ad}^c$ we expect formation of a $S=1$ state in
the regime when $\langle n \rangle \sim 2$. Since in this case $2S>K$
where $K=1$ is the number of channels, this systems falls into a class
where the spin on the DQD is not fully compensated by the conduction
electrons \cite{qpt,qptmultilevel}. Results at finite $J_{ad}$,
presented in Fig~\ref{fig6}, can be divided into two groups. For
$J_{ad}<J_{ad}^c$ the main effect of increasing $J_{ad}$ is seen as
expansion of the $S=1/2$ Kondo regime
\index{Kondo regime} where $\langle n\rangle \sim
1$. For $J_{ad}>J_{ad}^c$, two Kondo plateaus appear, one associated
with $S=1/2$ Kondo regime and the other with $S=1$ Kondo regime. In
both cases conductance reaches unitary limit, while in the transition
regime it drops to zero.

\section{Conclusions}
In this work  we have explored different regimes of the side-coupled
DQD. Conclusions can be summarized as follows: a) when quantum dots
\index{quantum dots}
are {\it strongly coupled}, wide regions of enhanced, nearly unitary
conductance exist due to the underlying Kondo physics.  Analytical
estimates for their positions, widths, as well as for the
corresponding Kondo temperatures, are given and numerically
verified. { When two electrons occupy DQD, conductance is zero due to
formation of the spin-singlet state which is effectively decoupled
from the leads. }  b) In the limit when quantum dots are {\it weakly
coupled} \index{quantum dots!weakly coupled}
the Fano resonance appears in the valence fluctuation regime. Its
width is enhanced as a consequence of interactions which should
facilitate experimental observation. Unitary conductance exists when
two electrons occupy DQD due to a 
two-stage Kondo effect\index{Kondo effect!two-stage} as long as
the temperature of the system is well below $T_K$ and above $T_K^0$.
The experimental signature of the two-stage Kondo effect in weakly
coupled regime should materialize through the inter-dot coupling
sensitive width of the enhanced conductance vs. gate
voltage.

When intra-dot ferromagnetic coupling exceeds a critical value, there
is a phase transition from a spin-singlet, non-conducting regime to
$S=1$ Kondo regime where the spin on the DQD is under-screened
but nevertheless the conductivity reaches the unitary limit.

\acknowledgements

Authors acknowledge useful discussions with A. Ram\v sak.  We also
acknowledge the financial support of the Slovenian Research Agency
under grant P1-0044.

\bibliographystyle{%
nato%
}
\bibliography{bonca-neu}

\printindex

\end{document}